\documentclass[10pt,a4paper]{article}
\usepackage{jheppub_kim}
\usepackage{pdflscape}
\usepackage{amsmath}
\usepackage{amssymb}
\usepackage{dcolumn}
\usepackage{bm}
\usepackage{color}
\usepackage{epsfig}
\usepackage{amsfonts}
\usepackage{graphicx}
\usepackage{subfigure}
\usepackage{dcolumn}

\usepackage{hyperref}

\begin{document}
\title{de Sitter Swampland Conjecture in String Field Inflation}
\author[a, d]{J. Sadeghi,}
\author[b]{B. Pourhassan,}
\author[b]{S. Noori Gashti,}
\author[c]{\.{I}. Sakall{\i},}
\author[a]{M. R. Alipour}
\affiliation[a] {Department of Physics, Faculty of Basic Sciences, University of Mazandaran P. O. Box 47416-95447, Babolsar, Iran.}
\affiliation[b] {School of Physics, Damghan University, Damghan, 3671641167, Iran.}
\affiliation[c] {Physics Department, Eastern Mediterranean University, Famagusta 99628, North Cyprus via Mersin 10, Turkey.}
\affiliation[d] {Canadian Quantum Research Center 204-3002 32 Ave Vernon, BC V1T 2L7 Canada.}

\emailAdd{pouriya@ipm.ir}
\emailAdd{b.pourhassan@du.ac.ir}
\emailAdd{saeed.noorigashti@stu.umz.ac.ir}
\emailAdd{izzet.sakalli@emu.edu.tr}
\emailAdd{mr.alipour@stu.umz.ac.ir}

\abstract{In this paper, we study a particular type of inflation by using non-local Friedman equations that are somehow derived from the zero levels of string field theory and express a tachyonic action. Then, we challenge it by further refining de Sitter (dS) swampland conjecture (FRdSSC) monitoring. Therefore, we investigate some quantities, such as potential and Hubble parameters. We also consider slow-roll parameters to examine quantities such as the scalar spectrum index and the tensor-to-scalar ratio. Using straightforward calculations, we investigate this model from the swampland conjecture perspective in terms of the cosmological parameters, i.e., ($n_s$), ($r$), and observable data such as Planck 2018, by constructing some structures such as $(c_1,2-n_s)$ and $(c_1,2-r_s)$. Then, we make a new restriction for this conjecture as $c_12c_22$ and get a limit for this model in the range $c0.0942$. We find this inflationary model is strongly in tension with the dS swampland conjecture (dSSC), i.e., $c_1=c_2 \neq \mathcal{O}(1)$. So, we shall challenge it with the FRdSSC, which has some free parameters, viz., $a,b>0$, $a+b=1$, and $q>2$. By setting these parameters, we examine the compatibility of the mentioned conjecture with this inflationary model. Finally, we infer from this string field inflation model that it satisfies the FRdSSC with the constraint of its free parameters $a$, $b$, and $q$.}


\maketitle

\section{Introduction}

In string field theory, strings are considered to be the fundamental building blocks of matter and are treated as extended objects rather than point particles. Second quantization is used to describe the behavior of these strings in terms of the creation and annihilation of particles, and to understand the interactions between them. Therefore, the relation between string field theory and second quantization is that second quantization is a tool used to study and understand the behavior of strings in string field theory. It allows for the description of the many-particle interactions and dynamics of strings, and helps to shed light on the fundamental nature of these objects and their role in the universe. String field theory has been studied at different levels and from the point of view of perturbative, which has also brought challenging results \cite{1,2}. While it is true that string field theory primarily uses perturbative methods, it is still possible to obtain non-perturbative information from this theory. For example, researchers have used the second quantized action to calculate off-shell amplitudes in string field theory. Additionally, string field theory has been applied to gain insights into string theory itself, making it an exciting area of study \cite{3,4,5,6}. String field theory can sometimes have instabilities, such as tachyonic instability. These instabilities can be examined using solutions, such as tachyonic condensation, to evaluate the stability of these cases \cite{7,8}. Using tachyonic potentials and unstable branes, the study of the D-brane/anti-D-brane system in string theory can be conducted within the context of a linear dilaton background \cite{Clement:2002mb,Sakalli:2013yha,Sakalli:2011nz,Sakalli:2010yy,Sakalli:2016fif,Sakalli:2014wja}.\\

In summary, other applications and studies that have been done in string field theory include the analysis of tachyon condensation with a large field\cite{9,10,11,12} and the study of noncommutative geometry\cite{13,14,15}. An essential point in string field theory is the existence of infinite fields. Besides, the actions for each field also have nonlocal sentences. In some studies, these nonlocal actions are used to investigate spontaneous symmetry breaking in the associated theories \cite{16,17,18}. This paper uses a nonlocal tachyonic approach to study inflation, which is often used in string field theory to create cosmological solutions \cite{19}. The equations of motion for such a structure are derived from the Hamilton-Jacobi formalism. Furthermore, the Friedmann's equations derived from nonlocal tachyonic fields are obtained and used in string theory \cite{20,21}. It is worth noting that one of the interesting aspects of string field theory is that it offers various nonlocal cosmological models that have been explored by scientists in recent times for their potential use in cosmological studies \cite{Kumar:2018chy}.\\

In general, nonlocal solutions derived from string theory can provide non-perturbation modifications for cosmological solutions. This nonlocality in cosmological solutions creates an acceleration expansion for the universe, which can have interesting cosmological applications. By applying the dSSC, we explore the potential cosmological implications of using this model to describe inflation \cite{22,23}. Moreover, de Sitter holography is also important in the inflationary scenario \cite{suss}. Therefore, the main focus of this article is to assess the compatibility of this inflation model with observational data by applying the de Sitter swampland conjecture and analyzing the results. In the literature, the inflation resolves a series of problems such as flatness, monopole, and horizon problems \cite{24,25,26,27,28}, in which it was described that how the concept of inflation can help us to explain the fluctuations observed in the temperature of the cosmic microwave background (CMB) and the formation of large structures in the universe. A scalar field is often associated with inflation and the process of inflation involves studying the evolution of the universe from its inflationary phase to its current state of expansion. Researchers have also examined various structures within the context of inflation, including effective low-energy theories, which have produced promising results \cite{29,30,31,32,33,34,35,36}. String theory also offers a group of inflation models that have significant implications for the study of the cosmos \cite{37,38,39,40}.\\

In the study of inflationary cosmology, it is common to observe distinct phases of development, such as the beginning stage of accelerated expansion and the final phase of reheating \cite{41,42,43}. There are two types of inflation: warm and cold. In warm inflation, the presence of an inflation field in a thermal bath is possible \cite{44,45,46,47}. As stated before, inflation has also been studied in string theory \cite{48,49,50,51,52}, including the fibre inflationary cosmology of type IIB string theory \cite{53,54} and string gas cosmology \cite{55,56}. Inflation with tachyon potential has even been studied in string theory \cite{57,58}. Moreover, the researchers have recently considered the swampland program and weak gravity conjecture, and a lot of work has been done concerning these conjectures. Both the cosmology of the weak gravity conjecture and the theories of swampland and landscape have been thoroughly described in terms of their applications. The weak gravity conjecture states that gravity is the weakest force in theories that are coupled to gravity, and swampland and landscape are sets of theories that are either incompatible or compatible with quantum gravity, respectively. In this regard, we refer the reader to check the following references \cite{59,60,61,62,63,64,65,66,67,68,69,70,71,72,73,74,75,76,77,78,79,80,81,a,b,c,d,e,JHAP} to learn more about those concepts and get acquainted with their cosmological applications.\\

In this study, we propose a specific form of inflation model by utilizing non-local Friedman equations and the perspective of swampland conjectures that are derived from the zero levels of string field theory and involve a tachyonic action. We shall present the FRdSSC for the inflation model and evaluate whether the model aligns with the swampland conjectures based on the results. This paper is organized as follows: In Sec. \ref{sec2}, we briefly explain the non-local string field theory and its corresponding equations of motion.
In Sec. \ref{sec3}, we investigate some quantities, such as tachyonic potential and Hubble parameters. We also consider slow-roll parameters to examine the quantities such as the scalar spectrum index and the tensor-to-scalar ratio. Using straightforward calculations, we investigate this model from the swampland conjecture perspective in terms of the cosmological parameters, i.e. $n_{s}$, $r$, and observable data such as Planck 2018. We discuss the swampland conditions for this inflation model by plotting some figures. We find the model is strongly in tension with the dSSC. So, we challenge this inflation model with the FRdSSC in Sec. \ref{sec4}. Finally, in Sec. \ref{sec5}, according to the results obtained, we draw our conclusions by discussing the compatibility or incompatibility of the prescribed inflation model with the swampland conjectures.

\section{The model overview} \label{sec2}

In this section, we will explore the concept of string field inflation (SFI) and its potential use in constructing a cosmological model. We will also delve into related ideas surrounding this theory, which is based on second quantized strings and can be expressed through a string field. One key feature of this string field is its ability to serve as a sum of string amplitudes in different states. As previously mentioned, the zero levels of the string field correspond to tachyonic fields, and this theory can be used to calculate non-local actions for these zero-level states. In addition, this theory has the ability to address gravitational singularities and can be studied as a nonlinear model in curved space-time, allowing for the development of cosmological solutions through the use of non-local actions for the zero-state modes in this curved space-time \cite{20,21,82,83,84,85,86,87,88}, whose action is given by
\begin{equation}\label{1}
S=\int d^{4}x\sqrt{-g}\bigg(\frac{M_{pl}^{2}}{2}R+\frac{1}{g \alpha^{'2}}\bigg[\frac{\alpha^{'}\xi{^2}}{2}\phi\Box\phi+\frac{1}{2}\phi^{2}-V\big(e^{\alpha^{'}k_{n}\Box}\phi\big)\bigg]\bigg),
\end{equation}
where $g$, $M_{pl}$, $\phi$, $k_{n}$, and V are the determinant of the metric $g_{\mu\nu}$, the Planck mass, the scalar field, the non-local coupling constant and the potential, respectively. Moreover, $\Box=\nabla_{\mu}\nabla^{\mu}=\partial_{\mu}\frac{(\sqrt{-g}g^{\mu\nu}\partial_{\nu})}{\sqrt{-g}}$. It is clear from action \eqref{1} that it is coupled with gravity. Such a connection would be useful for examining higher-level string field theory modes. Hence, this non-local action belomgs to a particular form, which is related to a specific string theory type. This action can also be expressed in the following form \cite{20,21}.
\begin{equation}\label{2}
S=\int d^{4}x\sqrt{-g}\bigg(\frac{\widehat{M_{pl}^{2}}}{2}R+\frac{\xi^{2}}{2}\phi\Box\phi+\frac{1}{2}\phi^{2}-V(\Phi)\bigg),
\end{equation}
where $\phi$ is a dimensionless scalar field, and $\Phi$ is a effective scalar field, i.e., $\Phi=e^{k\Box}\phi$ also $\widehat{M_{pl}^{2}}$=$g\alpha^{'}M_{pl}^{2}$ and $ k=\alpha^{'}k_{n}$. Also, we shall consider the following ansatz for the universe metric:
\begin{equation}\label{3}
ds^{2}=-dt^{2}+a^{2}(t)(dx^{2}+dy^{2}+dz^{2}).
\end{equation}
Taking account of Eq. \eqref{3} with two scalar fields $\phi$ and $\Phi$, one can obtain the following equation of motion:
\begin{equation}\label{4}
(1-\xi^{2}\mathcal{D}^{2})e^{2k\mathcal{D}^{2}}\Phi=V'(\Phi),
\end{equation}
where $\mathcal{D}^{2}=-\Box=\partial_{t}^{2}+3H(t)\partial_{t}$ and $H(t)=-\frac{\dot{a}(t)}{a(t)}$  is the Hubble parameter as well as $\frac{d\Phi}{d\phi}=e^{-k\mathcal{D}^{2}}$  and primes denote the differentiation with respect to $\Phi$. According to Eq. \eqref{4}, the value of $\xi$ reads: $\xi^{2}=-\frac{1}{4}\log(4/3\sqrt{3})$. Different values of this variable can always be used to obtain different cosmological parameters. Using Eqs. \eqref{2} and \eqref{3}, one can obtain the Friedmann equations as follows \cite{20,21,89}:

\begin{equation}\label{5}
3H^{2}=\frac{1}{\widehat{M_{pl}^{2}}}\mathcal{E}_{tot},\hspace{1cm}2\dot{H}+3H^{2}=-\frac{1}{\widehat{M_{pl}^{2}}}\mathcal{P}_{tot},
\end{equation}

where the index $"tot"$ represents the sum of local and non-local contributions of energy density and pressure. With respect to string field theory \cite{21,89}, we now have

\begin{equation}\label{6}
\mathcal{E}_{tot}=\mathcal{E}_{k}+\mathcal{E}_{p}+\mathcal{\widetilde{E}}_{1}+\mathcal{\widetilde{E}}_{2},\hspace{1cm}\mathcal{P}_{tot}=\mathcal{E}_{k}-\mathcal{E}_{p}-\mathcal{\widetilde{E}}_{1}+\mathcal{\widetilde{E}}_{2},
\end{equation}
where,
\begin{equation}\label{7}
\mathcal{E}_{k}=\xi^{2}\frac{(\partial_{t}\phi)^{2}}{2},
\end{equation}
\begin{equation}\label{8}
\mathcal{E}_{p}=-\phi^{2}+\frac{V(\Phi)}{2}
\end{equation}
\begin{equation}\label{9}
\mathcal{\widetilde{E}}_{1}=k\int_{0}^{1}d\varrho\big(e^{-k\varrho\mathcal{D}^{2}}V^{'}(\Phi)\big)\big(\mathcal{D}^{2}e^{k\varrho\mathcal{D}^{2}}\Phi\big),
\end{equation}
\begin{equation}\label{10}
\mathcal{\widetilde{E}}_{2}=-k\int_{0}^{1}d\varrho\big(\partial_{t}e^{-k\varrho\mathcal{D}^{2}}V^{'}(\Phi)\big)\big(\partial_{t}e^{k\varrho\mathcal{D}^{2}}\Phi\big),
\end{equation}

Concerning Eqs. \eqref{4}, \eqref{5}, \eqref{9}, and \eqref{10}, one can obtain

\begin{equation}\label{11}
\dot{H}=-\frac{1}{2\widehat{M}_{pl}^{2}}(\mathcal{P}_{tot}+\mathcal{E}_{tot}).
\end{equation}

Also, with regard to Eqs. \eqref{6} and \eqref{11}, and the above concept, one can find out

\begin{equation}\label{12}
H=-\frac{1}{\widehat{M}_{pl}^{2}}\int_{0}^{t}dt^{'}\bigg(\frac{\xi^{2}}{2}(\partial_{t'}\phi)^{2}-k\int_{0}^{1}d\varrho(\partial_{t'}(-\xi^{2}\mathcal{D}^{2}+1)e^{(2-\varrho)k\mathcal{D}^{2}}\Phi)\big(\partial_{t'}e^{\varrho k\mathcal{D}^{2}}\Phi\big)\bigg).
\end{equation}

To solve the above equation, we are faced with a non-local operator that can be solved using \cite{21}. Considering an ansatz to solve the problem and get an exact solution for the Hubble parameter, and in the sequel by using the Klein-Gordon equation \eqref{4}, this model's correct form of potential can be derived. Therefore, one can get the following conditions according to Ref. \cite{21}:

\begin{equation}\label{13}
\frac{\partial}{\partial \varrho}F(\varrho,t)= \mathcal{D}^{2}F(\varrho,t),\hspace{1cm}F(0,t)=F(t),\hspace{1cm}F(\varrho,\pm\infty)=F(\pm\infty).
\end{equation}

In reference to the above equation, one finds $-\infty<t<+\infty$ and $\varrho\geq 0$. At this stage, we consider a similar process as applied in \cite{21} and assume the following ansatz. We know that the exponential operator acts in such a way that $e^{k\varrho\mathcal{D}^{2}}F(t)=F(\varrho,t)$, which is unique for a class of Hubble parameters. So,

\begin{equation}\label{14}
F(\varrho,t)=\Phi(\varrho,t)=\alpha^{n}e^{\beta\varrho}t^{nq},
\end{equation}

where $\alpha$, $\beta$, $q$, and $n$ are the model parameters, by employing ansatz Eq. \eqref{14}, we solve Eq. \eqref{12} for the Hubble parameter. So we will have,

\begin{equation}\label{15}
H(t)=\frac{nq\alpha^{n}e^{\beta})^{2}(2k(\xi-\beta)+\xi^{2})}{2\widehat{M}_{pl}^{2}(1-2nq)}\times t^{2nq-1}.
\end{equation}

Furthermore, when we use Eqs. \eqref{4} and \eqref{15}, the potential is obtained as follows

\begin{equation}\label{16}
\begin{split}
V(t)=&\frac{9e^{2\beta}\alpha^{3n}\big(n^{2}q^{2}\alpha^{n}e^{\beta}\xi\big)^{2}\big(2k(\xi^{2}-\beta)+\xi^{2}\big)}{3\mathcal{M}_{pl}^{2}(5nq-2)}\times t^{5nq-2}\\
&+\frac{e^{2\beta}\alpha^{3n}(2nq-1)}{3}\bigg(\frac{6n^{2}q^{2}\xi^{2}(1-2nq)}{3nq-2}\times t^{-2+3nq}+t^{3nq}\bigg).\\
\end{split}
\end{equation}

One can use these non-local exact solutions obtained from string field theory to analyze inflation constraints in string field theory. Cosmological parameters such as slow-roll non-local string field inflation can be obtained. Therefore, we have

\begin{equation}\label{17}
\epsilon_{1}=-\frac{\dot{H}}{H^{2}},\hspace{1cm}\epsilon_{2}=-\frac{\dot{\epsilon}_{1}}{H\epsilon_{1}}.
\end{equation}

Also, the scalar spectral index and the tensor-to-scalar ratio are given by

\begin{equation}\label{18}
n_{s}-1=2\eta-6\epsilon,\hspace{1cm}r=16\epsilon.
\end{equation}

According to the definition of ansatz in Eq. \eqref{14}, where the effective scalar field is defined as $t(\Phi)=(\frac{\Phi}{\alpha})^{\frac{1}{n}}$ and using the descriptions of the slow-roll parameters in Eq. \eqref{17}, we can obtain any of the parameters as potential and Hubble and slow-roll parameters in terms of the effective scalar field. So, we have

\begin{equation}\label{19}
\epsilon=\frac{\widehat{M}_{pl}^{2}}{2}(\frac{V'(\Phi)}{V(\Phi)})^{2},\hspace{1cm}\eta=\widehat{M}_{pl}^{2}\frac{V''(\Phi)}{V(\Phi)}.
\end{equation}
In the next section, we shall study the effect of swampland conjecture on the inflation due to string field theory.

\section{SDC \& SFI} \label{sec3}
In this section, we will examine the inflation that was discussed in the previous section from the perspective of the swampland conjectures. We will calculate the potential and Hubble parameter for the effective scalar field based on ansatz \eqref{14} in the context of string field theory. We calculate the slow-roll parameters in order to analyze the scalar spectrum index and the tensor-to-scalar ratio. Through simple calculations, we study this model in relation to the swampland conjecture by considering the cosmological parameters, like ($n_{s}$) and ($r$), as well as data from the Planck 2018 observation. In this study, we present visual representations of the swampland conditions for the inflation model being examined. Based on these findings, we will determine whether the model is in accordance or conflict (compatibility/incompatibility) with the swampland conjectures. Despite efforts to differentiate between the landscape and swampland, there is still a lack of understanding of how to measure or detect gravity anomalies in these areas.

Currently, we only have a few theories or ideas about the principles of quantum gravity. These theories, known as swampland conjectures, have not yet been fully proven. However, in recent years, there has been a significant amount of evidence gathered for some of these conjectures, which are now widely accepted within the string theory community. One of the most interesting aspects of this research is the connections that are being discovered between these different conjectures. This discovery is exciting because it could potentially mean that these theories are different expressions of the same fundamental principles of quantum gravity that are gradually coming to light. The goal is for all the research and theories related to quantum gravity to eventually be integrated and contribute to a greater understanding of the concept. Developing hypotheses is just the beginning of the swampland program \cite{59,60,61,62,63,64,65,66,67,68,69,70,71,72,73,74,75,76,77,78,79,80,81,a,b,c,d,e}. However, there is still much work to be done before a claim can be confidently reached. Initially, we analyze the data we possess from string theory or black hole physics to discover universal patterns and create a framework. After that, a significant amount of effort is required to verify and potentially adjust these assumptions through experimentation, typically utilizing string theory or AdS/CFT as a stable quantum gravity framework for precise testing of the conjecture \cite{59,60,61,62,63,64,65,66,67,68,69,70,71,72,73,74,75,76,77,78,79,80,81,a,b,c,d,e}. Therefore, it is necessary to not only have a logical basis for a conjecture, but also to be able to explain the physical mechanisms behind it and what could potentially go wrong. This can be done through the use of black hole physics with established principles or another model-agnostic method. Combining these methods with testing of string theory can provide strong support for the conjecture. Examining the practical implications of these ideas is of great interest in the fields of particle physics and cosmology \cite{59,60,61,62,63,64,65,66,67,68,69,70,71,72,73,74,75,76,77,78,79,80,81,a,b,c,d,e}. On the other hand, dSSC is still an open question whether the vacuum string theory accepts the dS or not. There is no complete top-down structure of dS in a suitable regime of string theory. Although there is no specific case that is universally considered to be forbidden in string theory, finding a way to reconcile quantum gravity with the cosmic expansion of our universe is a fundamental and challenging goal. This is because of the potential consequences for our understanding of the universe, as well as the difficulties involved in constructing a dS vacuum within the framework of string theory. As a result, the swampland conjecture has been proposed to study the compatibility of different vacuums with quantum gravity, taking into account the connections with other swampland conjectures. In conclusion, we mention a conjecture that restricts the scalar potential in dS space. The dSSC and FRdSSC, which are part of the swampland program used to study cosmology, propose that effective theories of quantum gravity located in the landscape must meet at least one of the following conditions \cite{73}:
\begin{equation}\label{20}
|\nabla V|\geq\frac{c_{1}}{M_{p}}V, \hspace{12pt} min(\nabla_{i}\nabla_{j}V)\leq -\frac{c_{2}}{M_{pl}^{2}}V.
\end{equation}

The above equations for the $V>0$ can be rewritten in terms of the slow-roll parameters as follows:
\begin{equation*}\label{3}
\sqrt{2\epsilon_{V}}\geq c_{1} ,\hspace{12pt}  or \hspace{12pt}\eta_{V}\leq -c_{2},
\end{equation*}
where $c_1$ and $c_2$ are both positive constants and they have an order of one, i.e., $c_1=c_2=\mathcal{O}(1)$. Now, according to the ansatz of Eq. \eqref{14} and the explanations of the previous section, Eqs. \eqref{15} and \eqref{16} yield the Hubble parameter:

\begin{equation}\label{21}
H(\Phi)=\frac{\alpha^{2n}e^{2\beta}n^{2}q^{2}(-2k\beta+(1+2k)\xi^{2})\big((\frac{\Phi}{\alpha})^{\frac{1}{nq}}\big)^{-1+2nq}}{2\widehat{M}_{pl}^{2}(1-2nq)},
\end{equation}

and the potential is given by

\begin{equation}\label{22}
\begin{split}
V(\Phi)=&\frac{3\alpha^{5n}e^{4\beta}n^{4}q^{4}\xi^{2}\big(-2k\beta+(1+2k)\xi^{2}\big)\big((\frac{\Phi}{\alpha})^{\frac{1}{nq}}\big)^{-2+5nq}}{\widehat{M}_{pl}^{2}(-2+5nq)}\\
&+\frac{1}{3}\alpha^{3n}e^{2\beta}(-1+2nq)\big((\frac{\Phi}{\alpha})^{\frac{1}{nq}}\big)^{3nq}\bigg(1+\frac{6n^{2}q^{2}(1-2nq)\xi^{2}(\frac{\Phi}{\alpha})^{\frac{-2}{nq}}}{-2+3nq}\bigg).
\end{split}
\end{equation}

To investigate the swampland conjectures of Eq. \eqref{20}, we need to take the first and second potential derivatives. Thus, according to Eq. \eqref{22}, we have

\begin{equation}\label{23}
\begin{split}
V'=&\frac{1}{\widehat{M}_{pl}^{2}\Phi}\times(\frac{\Phi}{\alpha})^{\frac{-2}{nq}}\bigg[3\alpha^{5n}e^{4\beta}n^{3}q^{3}\xi^{2}(-2k\beta+(1+2k)\xi^{2})\big((\frac{\Phi}{\alpha})^{\frac{1}{nq}}\big)^{5nq}\\
&+\alpha^{3n}e^{2\beta}(-1+2nq)\big((\frac{\Phi}{\alpha})^{\frac{1}{nq}}\big)^{3nq}\widehat{M}_{pl}^{2}\bigg(2nq(1-2nq)\xi^{2}+(\frac{\Phi}{\alpha})^{\frac{2}{nq}}\bigg)\bigg]\\
\end{split}
\end{equation}
\begin{equation}\label{24}
\begin{split}
V''(\Phi)=&\frac{1}{\widehat{M}_{pl}^{2}\Phi^{2}}2\alpha^{3n}e^{2\beta}(-1+2nq)(\frac{\Phi}{\alpha})^{\frac{-2}{nq}}\big((\frac{\Phi}{\alpha})^{\frac{1}{nq}}\big)^{3nq}\\
&\times\bigg[3\alpha^{2n}e^{2\beta}n^{2}q^{2}\xi^{2}(-2k\beta+(1+2k)\xi^{2})\big((\frac{\Phi}{\alpha})^{\frac{1}{nq}}\big)^{2nq}\\
&+\widehat{M}_{pl}^{2}\bigg(-2(-1+nq)(-1+2nq)\xi^{2}+(\frac{\Phi}{\alpha})^{\frac{2}{nq}}\bigg)\bigg].
\end{split}
\end{equation}

Now, by using Eqs. \eqref{20}, \eqref{22}, \eqref{23}, and \eqref{24}, the swampland conjectures for this string field theory potential are calculated as follows:

\begin{equation}\label{25}
\begin{split}
&\frac{1}{\widehat{M}_{pl}^{2}\Phi}\times(\frac{\Phi}{\alpha})^{\frac{-2}{nq}}\bigg[3\alpha^{5n}e^{4\beta}n^{3}q^{3}\xi^{2}(-2k\beta+(1+2k)\xi^{2})\big((\frac{\Phi}{\alpha})^{\frac{1}{nq}}\big)^{5nq}\\
&+\alpha^{3n}e^{2\beta}(-1+2nq)\big((\frac{\Phi}{\alpha})^{\frac{1}{nq}}\big)^{3nq}\widehat{M}_{pl}^{2}\bigg(2nq(1-2nq)\xi^{2}+(\frac{\Phi}{\alpha})^{\frac{2}{nq}}\bigg)\bigg]\\
&\bigg/\frac{1}{\widehat{M}_{pl}^{2}\Phi}\times(\frac{\Phi}{\alpha})^{\frac{-2}{nq}}\bigg[3\alpha^{5n}e^{4\beta}n^{3}q^{3}\xi^{2}(-2k\beta+(1+2k)\xi^{2})\big((\frac{\Phi}{\alpha})^{\frac{1}{nq}}\big)^{5nq}\\
&+\alpha^{3n}e^{2\beta}(-1+2nq)\big((\frac{\Phi}{\alpha})^{\frac{1}{nq}}\big)^{3nq}\widehat{M}_{pl}^{2}\bigg(2nq(1-2nq)\xi^{2}+(\frac{\Phi}{\alpha})^{\frac{2}{nq}}\bigg)\bigg]>c_{1},
\end{split}
\end{equation}

\begin{equation}\label{26}
\begin{split}
&\frac{1}{\widehat{M}_{pl}^{2}\Phi^{2}}2\alpha^{3n}e^{2\beta}(-1+2nq)(\frac{\Phi}{\alpha})^{\frac{-2}{nq}}\big((\frac{\Phi}{\alpha})^{\frac{1}{nq}}\big)^{3nq}\\
&\times\bigg[3\alpha^{2n}e^{2\beta}n^{2}q^{2}\xi^{2}(-2k\beta+(1+2k)\xi^{2})\big((\frac{\Phi}{\alpha})^{\frac{1}{nq}}\big)^{2nq}\\
&+\widehat{M}_{pl}^{2}\bigg(-2(-1+nq)(-1+2nq)\xi^{2}+(\frac{\Phi}{\alpha})^{\frac{2}{nq}}\bigg)\bigg]\\
&\bigg/\frac{1}{\widehat{M}_{pl}^{2}\Phi}\times(\frac{\Phi}{\alpha})^{\frac{-2}{nq}}\bigg[3\alpha^{5n}e^{4\beta}n^{3}q^{3}\xi^{2}(-2k\beta+(1+2k)\xi^{2})\big((\frac{\Phi}{\alpha})^{\frac{1}{nq}}\big)^{5nq}\\
&+\alpha^{3n}e^{2\beta}(-1+2nq)\big((\frac{\Phi}{\alpha})^{\frac{1}{nq}}\big)^{3nq}\widehat{M}_{pl}^{2}\bigg(2nq(1-2nq)\xi^{2}+(\frac{\Phi}{\alpha})^{\frac{2}{nq}}\bigg)\bigg]<-c_{2}.
\end{split}
\end{equation}

After obtaining the exact values for the swampland conjectures, the slow-roll parameters are now calculated according to Eq. \eqref{19} as the following:

\begin{equation}\label{27}
\begin{split}
&A=(\frac{\Phi}{\alpha})^{\frac{-4}{nq}}\bigg[3\alpha^{5n}e^{4\beta}n^{3}q^{3}\xi^{2}(-2k\beta+(1+2k)\xi^{2})\big((\frac{\Phi}{\alpha})^{\frac{1}{nq}}\big)^{5nq}\\
&+\alpha^{3n}e^{2\beta}(-1+2nq)\big((\frac{\Phi}{\alpha})^{\frac{1}{nq}}\big)^{3nq}\bigg(2nq(1-2nq)\xi^{2}+(\frac{\Phi}{\alpha})^{\frac{2}{nq}}\bigg)\bigg]^{2},\\
&B=\frac{1}{2\widehat{M}_{pl}^{4}\Phi^{2}}\times\bigg[\frac{3\alpha^{5n}e^{4\beta}n^{4}q^{4}\xi^{2}(-2k\beta+(1+2k)\xi^{2})\big((\frac{\Phi}{\alpha})^{\frac{1}{nq}}\big)^{-2+5nq}}{\widehat{M}_{pl}^{2}(-2+5nq)}\\
&\frac{1}{3}\alpha^{3n}e^{2\beta}(-1+2nq)\big((\frac{\Phi}{\alpha})^{\frac{1}{nq}}\big)^{3nq}\bigg(1+\frac{6n^{2}q^{2}(1-2nq)\xi^{2}(\frac{\Phi}{\alpha})^{\frac{-2}{nq}}}{-2+3nq}\bigg)\bigg]^{2},\\
&\epsilon=\frac{A}{B},
\end{split}
\end{equation}

\begin{equation}\label{28}
\begin{split}
&A=\bigg\{6(-1+2nq)(-2+3nq)(-2+5nq)\bigg(3\alpha^{2n}e^{2\beta}n^{2}q^{2}\xi^{2}(-2k\beta+(1+2nq)\xi^{2})\big((\frac{\Phi}{\alpha})^{\frac{1}{nq}}\big)^{2nq}\\
&+\widehat{M}_{pl}^{2}\bigg[-2(-1+nq)(-1+2nq)\xi^{2}+(\frac{\Phi}{\alpha})^{\frac{2}{nq}}\bigg]\bigg)\bigg\}\\
&B=\bigg\{\Phi^{2}\bigg[9\alpha^{2n}e^{2\beta}n^{4}q^{4}(-2+3nq)\xi^{2}(-2k\beta+(1+2k)\xi^{2})\big((\frac{\Phi}{\alpha})^{\frac{1}{nq}}\big)^{2nq}\\
&+\widehat{M}_{pl}^{2}(-1+2nq)(-2+5nq)\bigg(6n^{2}q^{2}(1-2nq)\xi^{2}+(-2+3nq)(\frac{\Phi}{\alpha})^{\frac{2}{nq}}\bigg)\bigg]\bigg\},\\
&\eta=\frac{A}{B}.
\end{split}
\end{equation}

By using Eqs. \eqref{27} and \eqref{28}, one can obtain the exact values of the scalar spectrum index $n_{s}$ and the tensor-to-scalar ratio ($r$), i.e., Eq. \eqref{18}, which are expressed as follows

\begin{equation}\label{29}
\begin{split}
&A=2\bigg\{6(-1+2nq)(-2+3nq)(-2+5nq)\bigg[3\alpha^{2n}e^{2\beta}n^{2}q^{2}\xi^{2}(-2k\beta+(1+2nq)\xi^{2})\big((\frac{\Phi}{\alpha})^{\frac{1}{nq}}\big)^{2nq}\\
&+\widehat{M}_{pl}^{2}\bigg(-2(-1+nq)(-1+2nq)\xi^{2}+(\frac{\Phi}{\alpha})^{\frac{2}{nq}}\bigg)\bigg]\bigg\},\\
&B=\bigg\{\Phi^{2}\bigg[9\alpha^{2n}e^{2\beta}n^{4}q^{4}(-2+3nq)\xi^{2}(-2k\beta+(1+2k)\xi^{2})\big((\frac{\Phi}{\alpha})^{\frac{1}{nq}}\big)^{2nq}\\
&+\widehat{M}_{pl}^{2}(-1+2nq)(-2+5nq)\bigg(6n^{2}q^{2}(1-2nq)\xi^{2}+(-2+3nq)(\frac{\Phi}{\alpha})^{\frac{2}{nq}}\bigg)\bigg]\bigg\},\\
&C=-6(\frac{\Phi}{\alpha})^{\frac{-4}{nq}}\bigg[3\alpha^{5n}e^{4\beta}n^{3}q^{3}\xi^{2}(-2k\beta+(1+2k)\xi^{2})\big((\frac{\Phi}{\alpha})^{\frac{1}{nq}}\big)^{5nq}\\
&+\alpha^{3n}e^{2\beta}(-1+2nq)\big((\frac{\Phi}{\alpha})^{\frac{1}{nq}}\big)^{3nq}\bigg(2nq(1-2nq)\xi^{2}+(\frac{\Phi}{\alpha})^{\frac{2}{nq}}\bigg)\bigg]^{2},\\
&D=\frac{1}{2\widehat{M}_{pl}^{4}\Phi^{2}}\times\bigg[\frac{3\alpha^{5n}e^{4\beta}n^{4}q^{4}\xi^{2}(-2k\beta+(1+2k)\xi^{2})\big((\frac{\Phi}{\alpha})^{\frac{1}{nq}}\big)^{-2+5nq}}{\widehat{M}_{pl}^{2}(-2+5nq)}\\
&\frac{1}{3}\alpha^{3n}e^{2\beta}(-1+2nq)\big((\frac{\Phi}{\alpha})^{\frac{1}{nq}}\big)^{3nq}\bigg(1+\frac{6n^{2}q^{2}(1-2nq)\xi^{2}(\frac{\Phi}{\alpha})^{\frac{-2}{nq}}}{-2+3nq}\bigg)\bigg]^{2},\\
&n_{s}-1=\frac{A}{B}+\frac{C}{D},
\end{split}
\end{equation}

\begin{equation}\label{30}
\begin{split}
&A=16(\frac{\Phi}{\alpha})^{\frac{-4}{nq}}\bigg[3\alpha^{5n}e^{4\beta}n^{3}q^{3}\xi^{2}(-2k\beta+(1+2k)\xi^{2})\big((\frac{\Phi}{\alpha})^{\frac{1}{nq}}big)^{5nq}\\
&+\alpha^{3n}e^{2\beta}(-1+2nq)\big((\frac{\Phi}{\alpha})^{\frac{1}{nq}}\big)^{3nq}\bigg(2nq(1-2nq)\xi^{2}+(\frac{\Phi}{\alpha})^{\frac{2}{nq}}\bigg)\bigg]^{2},\\
&B=\frac{1}{2\widehat{M}_{pl}^{4}\Phi^{2}}\times\bigg[\frac{3\alpha^{5n}e^{4\beta}n^{4}q^{4}\xi^{2}(-2k\beta+(1+2k)\xi^{2})\big((\frac{\Phi}{\alpha})^{\frac{1}{nq}}\big)^{-2+5nq}}{\widehat{M}_{pl}^{2}(-2+5nq)}\\
&\frac{1}{3}\alpha^{3n}e^{2\beta}(-1+2nq)\big((\frac{\Phi}{\alpha})^{\frac{1}{nq}}\big)^{3nq}\bigg(1+\frac{6n^{2}q^{2}(1-2nq)\xi^{2}(\frac{\Phi}{\alpha})^{\frac{-2}{nq}}}{-2+3nq}\bigg)\bigg]^{2},\\
&r=\frac{A}{B}.
\end{split}
\end{equation}
To determine the impact of the swampland conjectures on the scalar spectrum index $(n_{s})$ and the tensor-to-scalar ratio ($r$) based on observable data such as Planck 2018 \cite{90}, we will first invert Eqs. \eqref{29} and \eqref{30} using an effective scalar field. Next, we will substitute this result into Eqs.{25} and \eqref{26}. This allows us to calculate the swampland conjectures in terms of $c_{1}-n_{s}$ and $c_{2}-n_{s}$, as well as $c_{1}-r$ and $c_{1}-r$. By plotting the data from \cite{90,91} and examining the resulting figures, we can see how the swampland conjectures vary with respect to $n_{s}$ and $r$.
\begin{figure}[h!]
\begin{center}
\subfigure[]{
\includegraphics[height=5cm,width=5cm]{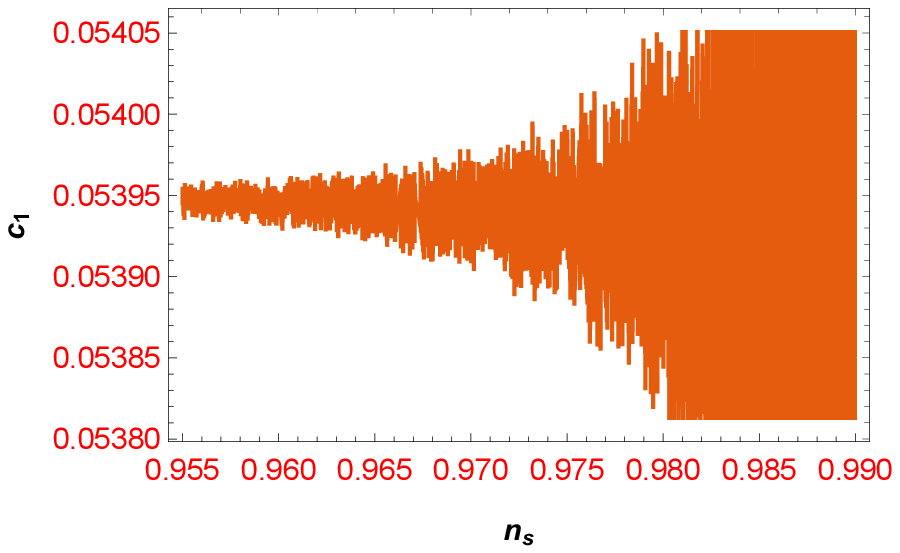}
\label{1a}}
\subfigure[]{
\includegraphics[height=5cm,width=5cm]{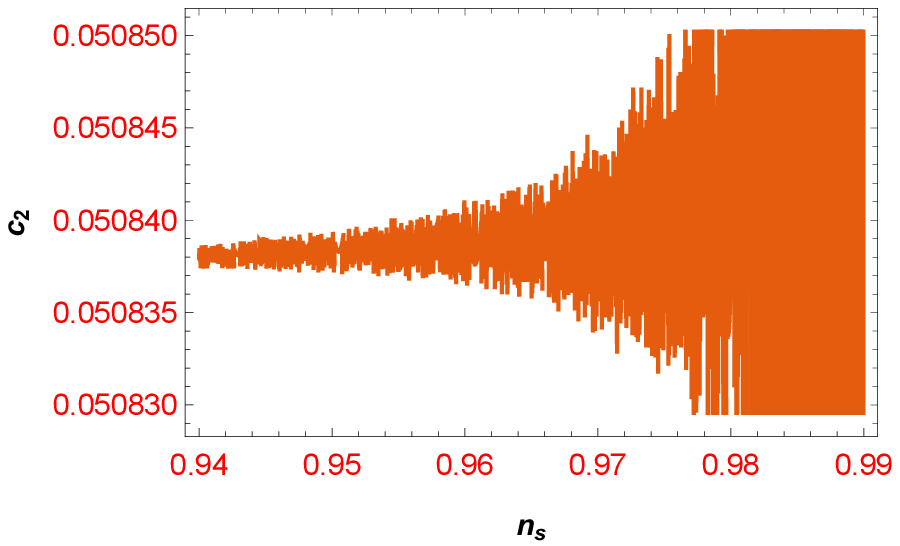}
\label{1b}}
\caption{\small{The plot of $c_{1}$ and $c_{2}$ in terms of $n_{s}$.}}
\label{fig1}
\end{center}
\end{figure}

\begin{figure}[h!]
\begin{center}
\subfigure[]{
\includegraphics[height=5cm,width=5cm]{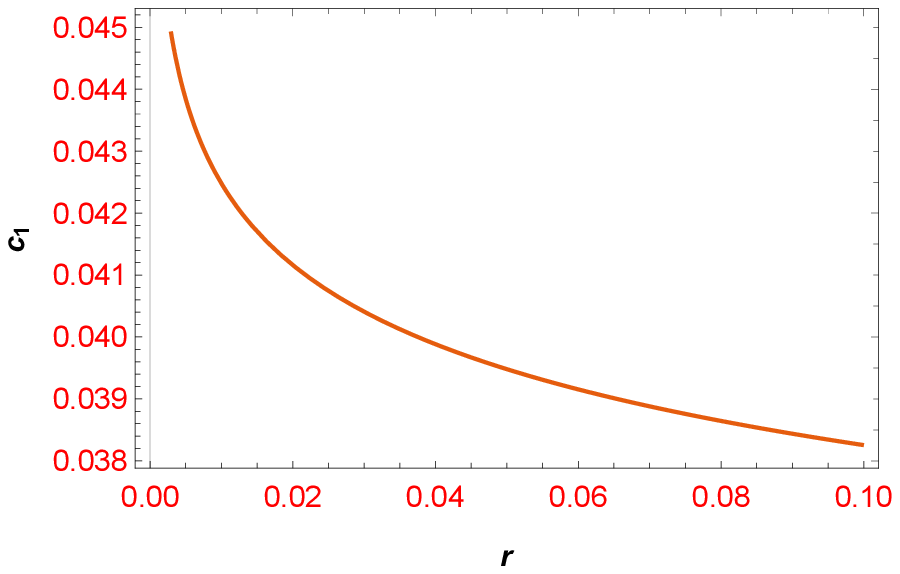}
\label{2a}}
\subfigure[]{
\includegraphics[height=5cm,width=5cm]{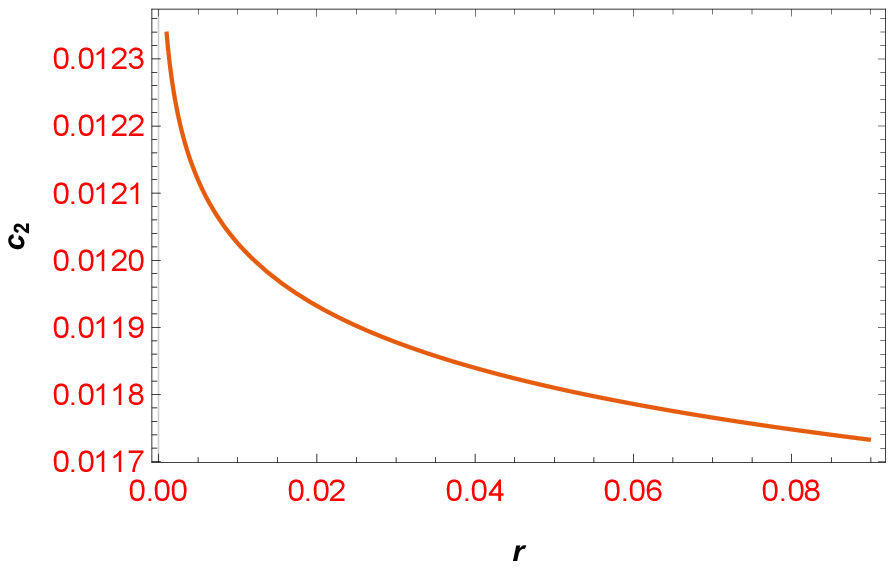}
\label{2b}}
\caption{\small{The plots of $c_{1}$ and $c_{2}$ versus $r$.}}
\label{fig2}
\end{center}
\end{figure}

\begin{figure}[h!]
 \begin{center}
 \subfigure[]{
 \includegraphics[height=6cm,width=6cm]{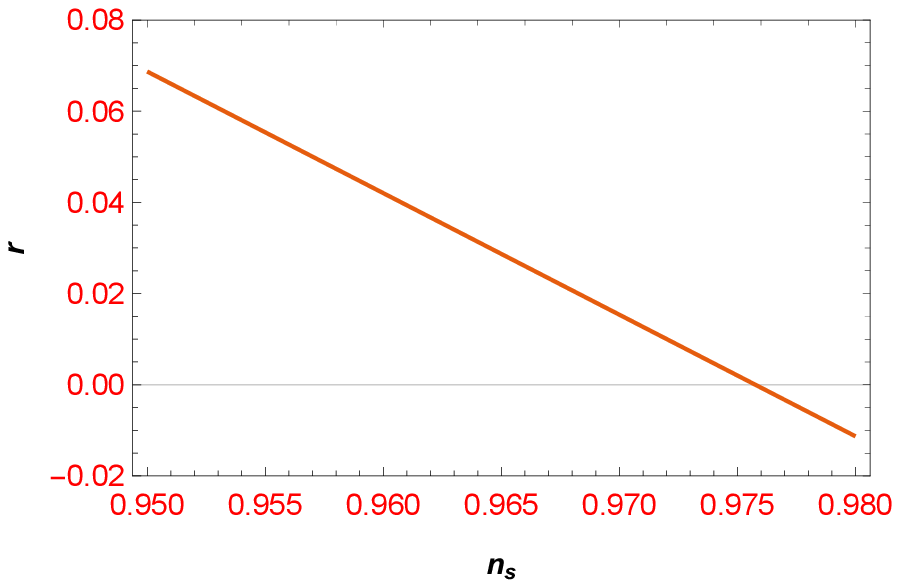}
}
 \caption{The plot of $r$ in term of $n_{s}$.}
 \label{fig3}
 \end{center}
 \end{figure}

The swampland conjectures are defined in terms of two cosmological parameters: the scalar spectrum index $(n_{s})$ and the tensor-to-scalar ratio ($r$). These parameters have acceptable values, and it is clear that the value of $c_{2}$ has fewer values than $c_{1}$. However, we can also look at how these two parameters change in relation to each other. The resulting values are similar to the observable values. We can plot these changes using Eqs. \eqref{29} and \eqref{30} and using recent observable data from \cite{90,91}. Figs. \ref{fig1}, \ref{fig2}, and \ref{fig3} show these changes.

\begin{figure}[h!]
 \begin{center}
 \subfigure[]{
 \includegraphics[height=6cm,width=6cm]{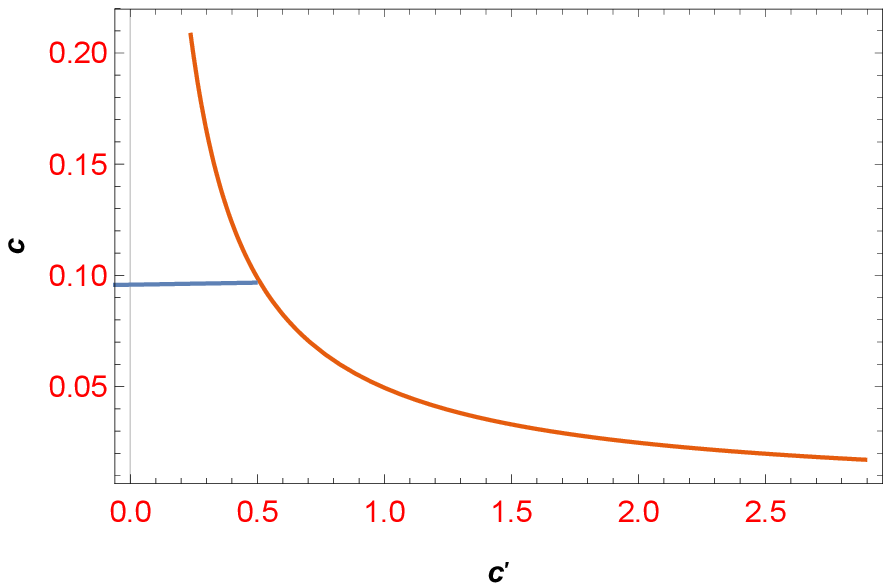}
 \label{4}}
 \caption{The plot of $c$ in term of $c'$.}
 \label{fig4}
 \end{center}
 \end{figure}

In this part, we use certain equations and previously discussed concepts to determine another restriction for the inflation model under consideration. We calculate the value of ($c$) and specify its range based on the limitations and available data. As illustrated in Fig. \ref{fig4}, according to the condition $c=f>(c_{1}^{2})(c_{2}^{2})$ and the available data, the limit of the swampland condition is $c<0.0942$. Taking into account the observable data, this constraint is presented below. It is also possible to test this inflation model using other conditions, such as FRdSSC, which will be examined in a separate section.

\section{FRdSSC \& SFI} \label{sec4}
In this part of the study, we will examine the potential for string field inflation by applying the modified dSSC, which has been put forth by Andriot and Roupec in their recent work \cite{92}. This conjecture posits that any effective low-energy theories that are compatible with quantum gravity must fulfill certain criteria at any location in space where the value of $V$ is greater than zero \cite{92}:

\begin{equation}\label{31}
\bigg(M_{p}\frac{|\nabla V|}{V}\bigg)^{q}-aM_{p}^{2}\frac{min(\nabla_{i}\nabla_{j}V)}{V}\geq b,
\end{equation}
where $a$, $b$, and $q$ are its free constant parameters that create a restriction for this conjecture. Additionally, it is important to note that the variables $a$, $b$, and $q$ are all positive and that $a$ and $b$ add up to 1. Using Eqs. \eqref{19} and \eqref{31}, we can express the FRdSSC in terms of slow-roll parameters.
\begin{equation}\label{32}
\big(2\epsilon \big)^{\frac{q}{2}}-a\eta \geq 1-a.
\end{equation}
In the following, we examine the FRdSSC for string field inflation according to $n_s$ and $r$ obtained in article \cite{91}.
According to Table I illustrated in article \cite{91}, we choose two values of ($n_s$ , $r$) and examine them. For example ($n_s=0.960136$ , $r=0.90595 \times 10^{-9}$) and ($n_s=0.944485$ , $r=2.30056 \times 10^{-9}$).
We represent the parameters $F_1$ and $F_2$, which can be helpful, as follows:
\begin{equation}\label{33}
F_1=\frac{|V'(\phi)|}{V(\phi)}=\sqrt{2\epsilon} \hspace{1.5cm}   F_2=\frac{V''(\phi)}{V(\phi)}=\eta.
\end{equation}
In this case, we rewrite Eq. \eqref{32} in terms of $F$ as follows
\begin{equation}\label{34}
(F_1)^{q}-a F_2 \geq 1-a
\end{equation}
Also, with a straightforward analysis, we can rewrite $F_1$ and $F_2$ in terms of tensor-to-scalar ratio $r$ and scalar spectrum index $n_s$, respectively, as follows:
\begin{equation}\label{35}
F_1=\sqrt{2\epsilon}=\sqrt{\frac{r}{8}}, \hspace{1.5cm}   F_2=\eta=\frac{1}{2}(n_s-1+\frac{3r}{8}).
\end{equation}
Now, we examine the inflation model with the FRdSSC with respect to ($n_s=0.960136$ , $r=0.90595 \times 10^{-9}$) for  SFI. By putting these values in Eq. \eqref{35}, we get:
\begin{equation}\label{36}
F_1=\sqrt{2\epsilon}=\sqrt{\frac{r}{8}}=1.06416\times 10^{-5} \hspace{1.5cm}   F_2=\eta=\frac{1}{2}(n_s-1+\frac{3r}{8})=-0.019932.
\end{equation}
Considering the FRdSSC, we find:
\begin{equation}\label{37}
c_{1}\leq 1.06416\times 10^{-5}, \hspace{1.5cm}
\hspace{1.5cm}
c_2 \leq 0.019932,
\end{equation}
where $c_{1}$ and $c_2$ do not belong to $\mathcal{O}(1)$. So, we inform that the SDC is not satisfied for this inflationary model. According to Eqs. \eqref{34} and \eqref{36}, we can obtain a FRdSSC for ($n_s=0.960136$ , $r=0.90595 \times 10^{-9}$) as follows,
\begin{equation}\label{38}
(1.06416\times 10^{-5})^q+ 0.019932 a \geq 1-a, \hspace{1cm} \text{or} \hspace{1cm}(1.06416\times 10^{-5})^q \geq 1- 1.019932 a.
\end{equation}
We can find $a$ to satisfy the condition
\begin{equation}\label{39}
\frac{1}{1.019932}[1-(1.06416\times 10^{-5})^q]\leq a <1 ,  \hspace{1cm} q>2.
\end{equation}
Therefore, according to Eq. \eqref{39} for all values of $q>2$, the FRdSSC is satisfied. For example, when $q=2.3$, we get $0.980458 \leq a < 1$ and by choosing $a=0.985758$ we will have  $b=1-a=0.014242$.\\
Now, we use the above calculations in Eq. \eqref{35} with ($n_s=0.944485$ and $r=2.30056 \times 10^{-9}$) and get
\begin{equation}\label{40}
F_1=\sqrt{2\epsilon}=\sqrt{\frac{r}{8}}=1.69579\times 10^{-5}, \hspace{1.5cm}   F_2=\eta=\frac{1}{2}(n_s-1+\frac{3r}{8})=-0.0277575.
\end{equation}
Hence, we find
\begin{equation}\label{41}
c_{1}\leq 1.69579\times 10^{-5}, \hspace{1.5cm}   c_2 \leq 0.0277575.
\end{equation}

Like the previous case, here $c_{1}$ and $c_2$ also do not belomg to $\mathcal{O}(1)$ \textit{viz} SDC is not satisfied for this inflationary model. We challenge it with FRdSSC.

\begin{equation}\label{42}
(1.69579\times 10^{-5})^q+ 0.0277575 a \geq 1-a, \hspace{1cm} or \hspace{1cm} (1.69579\times 10^{-5})^q \geq 1- 1.0277575 a.
\end{equation}
We can find $a$ to satisfy the condition
\begin{equation}\label{43}
\frac{1}{1.0277575}[1-(1.69579\times 10^{-5})^q]\leq a <1 ,  \hspace{1cm} q>2.
\end{equation}
According to the above equation, the FRdSSC will hold for all $q>2$ values. For example, when $q=2.3$ we obtain $0.972992 \leq a < 1$ and by choosing $a=0.980125$, we will have  $b=1-a=0.019875$. The above calculations conclude that the string field inflation model satisfies the FRdSSC.

\section{Conclusion} \label{sec5}
In this research paper, we analyzed a specific type of inflation using non-local Friedman equations that come from the zero levels of string field theory and involve a tachyonic action. We then tested it against the FRdSSC monitoring system. We looked at various quantities such as the potential and Hubble parameters, as well as slow-roll parameters such as the scalar spectrum index ($n_{s}$) and the tensor-to-scalar ratio ($r$). Using mathematical calculations, we examined this model in relation to the swampland conjecture and cosmological parameters, including $n_{s}$ and $r$, as well as observable data like the Planck 2018. We created structures like $(c_{1,2}-n_{s})$ and $(c_{1,2}-r_{s})$ to do so. We also established a new restriction for the conjecture, $c_{1}^{2}c_{2}^{2}$, which resulted in a limit for this model of $c<0.0942$. We found that this inflationary model was in strong tension with the dSSC, meaning $c_{1}=c_{2}\neq \mathcal{O}(1)$. We tested the compatibility of a certain conjecture with the FRdSSC inflationary model, which has adjustable parameters $a, b>0$ such that $a + b = 1$ and $q>2$. By adjusting these parameters, we found that this string field inflation model met the requirements of the FRdSSC when certain values were chosen for $q$, specifically $q=2.3$ resulted in the ranges $0.972992 \leq a < 1$ and $b=1-a=0.019875$. It would be interesting to explore in the future whether this inflationary model satisfies other swampland conjectures and whether it can accurately model the evolution of the universe. This will be in our near future to-do agenda.

\end{document}